\documentstyle[floats,aps,epsfig,prb]{revtex}

\begin{document}
\draft

\twocolumn[\hsize\textwidth\columnwidth\hsize\csname
@twocolumnfalse\endcsname

\title{Monoclinic and triclinic phases in higher-order Devonshire theory}
\author{David Vanderbilt and Morrel H. Cohen}
\address{Department of Physics and Astronomy, Rutgers University,
Piscataway, New Jersey 08854-8019}

\date{November 16, 2000}
\maketitle

\begin{abstract}
Devonshire theory provides a successful phenomenological
description of many cubic perovskite ferroelectrics such as
BaTiO$_3$ via a sixth-order expansion of the free energy in the
polar order parameter.  However, the recent discovery of a novel
monoclinic ferroelectric phase in the PZT system by Noheda {\it et
al.}\ (Appl.\ Phys.\ Lett.\ {\bf 74}, 2059 (1999))
poses a challenge to this theory.  Here, we confirm that the
sixth-order Devonshire theory cannot support a monoclinic phase,
and consider extensions of the theory to higher orders.  We show
that an eighth-order theory allows for three kinds of equilibrium
phases in which the polarization is confined not to a symmetry axis
but to a symmetry plane.  One of these phases provides a natural
description of the newly observed monoclinic phase. Moreover, the
theory makes testable predictions about the nature of the phase
boundaries between monoclinic, tetragonal, and rhombohedral
phases.  A ferroelectric phase of the lowest (triclinic) symmetry
type, in which the polarization is not constrained by symmetry,
does not emerge until the Devonshire theory is carried to twelfth
order.  A topological analysis of the critical points
of the free-energy surface facilitates the discussion of the phase
transition sequences.
\end{abstract}

\vskip1pc]

\section{Introduction}
\label{sec:intro}

Many of the most important and best studied ferroelectric materials
adopt the simple-cubic perovskite structure at high temperature
and undergo structural phase transitions to distorted ferroelectric 
structures at lower temperature.  Among the best known simple compounds
of this kind are BaTiO$_3$ and PbTiO$_3$.
Upon cooling, BaTiO$_3$ undergoes a sequence of ferroelectric
transitions, first from the cubic (C) to a tetragonal (T), then
to an orthorhombic (O), and finally to a rhombohedral (R) phase.
Passing through this sequence, the polarization $\bf P$
first vanishes in the C phase, and then becomes oriented in the [001],
[011], and [111] directions in the T, O, and R phases, respectively.
PbTiO$_3$ undergoes a single ferroelectric transition from the C
to T phase.

In a classic 1948 paper, Devonshire \cite{dev} was able to explain
the observed phases and phase transition sequence quite naturally
in terms of a phenomenological Landau-type expansion of the
free energy in terms of the ferroelectric order parameter $\bf P$.
Making use of cubic symmetry and truncating the expansion to sixth
order in $\bf P$, Devonshire was able to arrive at a simple model
with only a single temperature-dependent second-order coefficient, and
only three temperature-independent higher-order coefficients.
The polarization $\bf P$ is the primary order parameter, and the
crystallographic labels (T, R, etc.) refer to the distortions
induced by the polarization and the resulting strain.
Despite its simplicity, this model could successfully reproduce
the phase-transition sequence\cite{dev} and the piezoelectric
and other properties \cite{devmore} of BaTiO$_3$.  With a simple
modification of the anharmonic coefficients, the qualitative
behavior of PbTiO$_3$ could be equally well reproduced.

However, the material that is currently in most wide\-spread use for
piezoelectric transducer and related applications is the solid
solution PbZr$_{1-x}$Ti$_x$O$_3$, commonly known as PZT.
The standard understanding of the phase diagram of PZT has been
as follows.\cite{jaffe} PZT undergoes a transition from the simple
cubic C phase to a ferroelectric phase at a Curie temperature
$T_c$ that ranges from about 490$^\circ$C at $x$=1 to 230$^\circ$C at
$x$=0.  The transition occurs to the T phase for $x$ greater than
about 0.48, and to the R phase for smaller $x$.  At $T<150^\circ$C
or $x<0.1$, some more complex phases involving antiferroelectric (AFE)
or antiferrodistortive (AFD, i.e., involving rotations of oxygen
octahedra) displacements also occur.  The phase boundary between
the simple T and R phases, known as the morphotropic phase boundary
(MPB), is an almost vertical line in the $x$--$T$ plane at about
$x$=0.48.  Haun {\it et al.}\cite{haun} have successfully extended
the Devonshire model to the case of PZT by including the AFE and
AFD degrees of freedom in the phenomenological free energy, still
including only terms up to sixth order overall.  This model
successfully described the simple R--T transition that was
understood to occur below $T_c$ at the MPB.

As it happened, a surprise was in store.  By working with highly
purified and carefully prepared samples of PZT, Noheda and coworkers
have recently shown\cite{noh-apl,noh-prl,noh-prb,noh-prb2} that a sliver of
monoclinic (M) phase actually interposes itself between the R and T
phases in a very narrow composition range (of order 3-4\% in
$x$).  That is, at least below $\sim$100$^\circ$C, the transition is
first from T to M at an $x_{c2}$ between about 0.48 and 0.51 (depending
on temperature), and then from M to R at $x_{c1}\simeq0.47$, with
decreasing $x$.  The orientation of $\bf P$ is, respectively, along
[001], [$uuv$] ($u<v$), and [111] in the T, M, and R phases,
respectively.  The experiments have not yet clarified whether or not
a direct T--R transition occurs in the higher temperature range
100$^\circ$C$<T< T_c$, or whether the sliver of M phase instead persists
up to the Curie temperature $T_c\simeq370^\circ$C.  Neither the
Devonshire theory,\cite{dev} nor the modification of Haun {\it et
al.},\cite{haun} predicted the possible occurrence of the M phase.  On
the other hand, simulations based on a first-principles effective
Hamiltonian approach\cite{zhong} have very recently provided
confirmation of the existence of the M phase in just the observed
composition range.\cite{bellaiche}

We make a brief aside to establish notation.  In the C phase (space
group $Pm3m$), $\bf P=0$.  When $\bf P$ is constrained to a symmetry axis
lying along [001], [111], or [011], the resulting phase and space-group
labels become T ($P4mm$), R ($R3m$), or O ($Amm2$), respectively.
Similarly, the M phases arise when $\bf P$ is confined to a mirror
plane.  We can distinguish three cases: M$_{\rm C}$ ($Pm$), in
which $\bf P$ is along [$0uv$]; and M$_{\rm A}$ and M$_{\rm B}$
(both $Bm$, sometimes also denoted $Cm$) in which $\bf P$ lies
along [$uuv$], with $u<v$ and $u>v$, respectively.  The newly
observed phase of PZT is of type M$_{\rm A}$.\cite{noh-prb,noh-prb2}  If
$\bf P$ is unconstrained by symmetry, strain coupling leads to a
triclinic phase ($P1$).  This exhausts the possible reduced-symmetry
states of a cubic perovskite crystal generated by the emergence a
single non-zero ferroelectric order parameter, although other phases
can occur if AFE and AFD distortions are also present.
 
Relatively few examples are known of low-symmetry ferroelectrics in
which $\bf P$ is only constrained to a symmetry plane, or in which $\bf
P$ is unconstrained by symmetry.  The discovery of the $M_{\rm A}$
phase in a cubic perovskite is thus of considerable note, even aside
from the fact that it had been missed for so long and aside from the
potential importance of this phase for understanding the large
piezoelectric response in PZT.\cite{noh-prl}

The failure of the phenomenological theories of Devon\-shire \cite{dev}
and Haun {\it et al.} \cite{haun} to describe the existence of the
observed M$_{\rm A}$ phase raises an interesting question: What is the
simplest and most natural phenomenological model that {\it does}
predict such a monoclinic phase?  Recent work of Souza Filho {\it et
al.}\cite{filho} confirms that the sixth-order Devonshire expansion
does not allow for the occurrence of a monoclinic phase and suggests
instead a model in which the vanishing of a shear elastic constant at
the critical temperature drives the transition to the M$_{\rm A}$
phase.  However, this requires the introduction of an additional
instability that is unrelated to the ferroelectric one.
\cite{explan-filho}  That this is unnecessary is demonstrated by the
work of Bellaiche, Garc\'{\i}a and Vanderbilt.\cite{bellaiche}  In their
simulations, the shear moduli are taken as temperature-independent, and
yet the transition to the monoclinic phase still occurs, driven by the
tilting of the ferroelectric polarization away from the symmetry
axis.

Suspecting instead that the problem is simply related to the
truncation of the expansions of Devonshire\cite{dev} and Haun {\it
et al.}\cite{haun} to sixth order, we consider the addition of
terms of eighth and higher order to the Devonshire model.  The
following questions then arise.  At what order in the expansion do
monoclinic phases first appear in the
phase diagram?  And, for that matter, at what order do triclinic
phases first appear?

The phase diagram of such a model consists of fields in the parameter
space of the model (labeled C, T, O, R, M, etc.) within which the
order parameter has the specified symmetry (that is, the absolute
minimum of the
free energy occurs at a point in the order-parameter space of that
symmetry).  On a trajectory in the phase space which crosses a phase
boundary, that global minimum can change into a local minimum, a
saddle point, or a maximum, or simply disappear.  Understanding how
the set of critical or stationary points of the free energy (its
minima, saddle points, and maxima in the order-parameter space) varies
with the parameters of the model can thus be an important aid in
understanding the phase diagram. So, in addition to searches for the
global minima by direct computation, we use here a topological
analysis of the Devonshire theory to elucidate the critical-point
structure of the free energy and answer the above questions.

We show here that the simplest extension of the Devonshire theory,
to eighth order in the ferroelectric order parameter, naturally admits
all three kinds (M$_{\rm A}$, M$_{\rm B}$, M$_{\rm C}$) of
monoclinic phases.  Moreover, the eighth-order model makes specific
and testable predictions for the types of phase boundaries that can occur.
For example, it predicts that the R--M$_{\rm A}$ and T--M$_{\rm A}$
transitions should be of first and second order, respectively.
On the other hand, we find that the model has to be extended all
the way to {\it twelfth} order in order to describe a triclinic
ferroelectric phase.

We therefore suggest that the most natural explanation for the
occurrence of the M$_{\rm A}$ phase in PZT is that the free-energy
surface is unusually anharmonic in this material, such that the
eighth-order terms play an important role.  It is not difficult to
speculate why this might be the case.  First, PZT is a disordered
material; averaging over the chemical disorder, which plays the role of
a quenched random field, may tend to generate higher orders in the
phenomenological energy expansion.  Second, as will be discussed in
Sec.~\ref{sec:model}, there is considerable evidence that the
ferroelectric transitions in PZT have a strong order-disorder
character; mapping onto a displacive picture may then also tend to
generate higher-order terms.  And finally, of course, thermal
fluctuations and coupling to strain may play some role.  In any
case, we show below that once one accepts this simple hypothesis
of the importance of eighth-order terms in the free-energy expansion,
then the behavior of the monoclinic phase observed in experiments
\cite{noh-apl,noh-prb,noh-prb2} and simulations \cite{bellaiche} can be
understood quite naturally.

This paper is organized as follows.
Section~\ref{sec:form} establishes the notation used for the expansion
of the free energy, and reviews the symmetry considerations that
lead to a simplified form of this expansion.
The rules governing the numbers and types of stationary points
that may occur in the order-parameter space are reviewed in
Sec.~\ref{sec:topology}.  The behavior of the models
obtained by truncating the expansion at higher and higher order
(fourth, sixth, eighth, tenth, and twelfth) are then carefully elucidated
in Sec.~\ref{sec:results}.  Section~\ref{sec:model} gives a brief
discussion of a microscopic model in which the T and R phases can be
regarded as arising from fluctuations among neighboring,
symmetry-equivalent local M$_{\rm A}$ states.  Finally,
we conclude with a brief summary and discussion of future
prospects in Sec.~\ref{sec:sum}.

\section{Formalism}
\label{sec:form}

We consider the case of a structural phase transition governed by
a single continuous vector order parameter $\bf u$, such that the
free energy
\begin{equation}
F({\bf u},\sigma_i,T)=E({\bf u},\eta_i,S)-\sum_i\sigma_i\eta_i-TS
\label{eq:free}
\end{equation}
is symmetric with respect to operations of the cubic point group.
Here $\sigma_i$ and $\eta_i$ are the stress and strain tensors in
Voigt notation, and $T$ and $S$ are temperature and entropy, respectively.
We have in mind primarily the case in which $\bf u$ is the
ferroelectric polarization $\bf P$ in a member of the cubic perovskite
class, such as BaTiO$_3$ or PZT, but the use of the symbol $\bf u$
serves as a reminder that the formalism applies to a variety of other
cases as well.  For a crystal with stress-free boundary conditions at
some given temperature, we can expand
\begin{eqnarray}
F({\bf u}) && =C_{000} + C_{200}\,(u_x^2+u_y^2+u_z^2)
+ C_{400}\,(u_x^4+u_y^4+u_z^4)
\nonumber\\ &&
+ C_{220}\,(u_x^2u_y^2+u_x^2u_z^2+u_y^2u_z^2)
+ C_{600}\,(u_x^6+u_y^6+u_z^6)
\nonumber\\ &&
+ C_{420}\,(u_x^4[u_y^2+u_z^2]+u_y^4[u_x^2+u_z^2]+u_z^4[u_x^2+u_y^2])
\nonumber\\ &&
+ C_{222}\,u_x^2u_y^2u_z^2
+ ...,
\label{eq:expan}
\end{eqnarray}
where terms up to sixth order in $u$ have been written explicitly,
and the coefficients have been renormalized to subsume the couplings
to strain.

For the ferroelectric phases we
simplify matters further by focusing on the energy as a function of
the {\it orientation} of the vector order parameter and introducing the
function
\begin{equation}
G(\hat{\bf u})=\min_{\bf u\parallel\hat{u}}\,F({\bf u})
\;\;.
\label{eq:gfunc}
\end{equation}
Thus, $G(\hat{\bf u})$ represents the ground-state energy subject
to the constraint that the order parameter has given orientation.
In most cases of interest, it is reasonable to expect that
$G(\hat{\bf u})$ will be a smooth function of
$\hat{\bf u}$.\cite{explan-a} In this case, and suppressing the
uninteresting constant term in the expansion, it follows that
\begin{eqnarray}
G(\hat{\bf u})=&a_4g_4&+a_6g_6+a_8g_4^2 \nonumber\\
&+&a_{10}g_4g_6+a_{12}g_4^3 +a_{12}'g_6^2+... \;\;,
\label{eq:gexpan}
\end{eqnarray}
where
\begin{eqnarray}
g_4(\hat{\bf u}) &=& x^2y^2+x^2z^2+y^2z^2 \nonumber \\
g_6(\hat{\bf u}) &=& x^2y^2z^2
\label{eq:gfuncs}
\end{eqnarray}
and $\hat{\bf u}=(x,y,z)$ with $x^2+y^2+z^2=1$.  All independent,
symmetry-allowed terms up to twelfth order are explicitly given in
Eq.~(\ref{eq:gexpan}); higher orders will not be needed here.

The particular form of the cubic invariants appearing in
Eq.~(\ref{eq:gexpan}) is largely arbitrary.  For example, one could
use the ``kubic harmonics'' \cite{kubharm} instead.
In Sec.~\ref{sec:higher}, we shall make use of the expansion of
Eqs.~(\ref{eq:gexpan}-\ref{eq:gfuncs}) above.  However,
for the presentation of numerical results to be given in
Sec.~\ref{sec:eighth}, we find it more convenient instead to use
the expansion
\begin{equation}
G(\hat{\bf u})=c_4f_4+c_6f_6+c_8f_8+... \;\;,
\label{eq:fexpan}
\end{equation}
where
\begin{eqnarray}
f_4(\hat{\bf u})&=&12g_4 \nonumber\\
f_6(\hat{\bf u})&=&4g_4-36g_6 \nonumber\\
f_8(\hat{\bf u})&=&48g_4^2-12g_4-36g_6 \;\;.
\label{eq:ffuncs}
\end{eqnarray}
Aside from normalization constants, which are chosen to
make the range of each function roughly of order unity, this
choice can be uniquely defined by the following requirements.
(i) The invariant $f_n$ contains no terms of order higher than $n$.
(ii) All three invariants vanish identically for a ``tetragonal'' (T)
value of the order parameter, e.g., $\hat{\bf u}=(100)$.
(iii) $f_6$ and $f_8$ still vanish for a ``rhombohedral'' (R)
value of the order parameter, e.g., $(111)/\sqrt{3}$.
(iv) $f_8$ alone vanishes for an ``orthorhombic'' (O) value
of the order parameter, e.g., $(110)/\sqrt{2}$.  The values of these
functions evaluated at symmetry directions are summarized in Table
I for later reference.  Requirement (ii) just reflects an arbitrary
choice of zero; requirements (iii) and (iv) simplify some later
discussion.  For example, phase boundaries at which $G$(T), $G$(R),
and $G$(O) become degenerate in pairs are easily located
(see Sec.~\ref{sec:eighth}).

\begin{table}
\caption{Values of the cubic invariant functions defined
in Eqs.~(\ref{eq:gfuncs}) and (\ref{eq:ffuncs}), evaluated
at tetragonal (T), rhombohedral (R), and orthorhombic (O)
orientations of the order parameter $\hat{\bf u}$.}
\begin{tabular}{cccc}
$F(\hat{\bf u})$ & $F$(T) & $F$(R) & $F$(O) \\
\hline
$g_4$ & 0 & $1/3$ & $1/4$ \\
$g_6$ & 0 & $1/27$ & 0 \\
$f_4$ & 0 & 4 & 3 \\
$f_6$ & 0 & 0 & 1 \\
$f_8$ & 0 & 0 & 0 \\
\end{tabular}
\label{tab:funcvals}
\end{table}

A few comments about the transition from $F(\bf u)$ to $G(\hat{\bf u})$
in Eq.~(\ref{eq:gfunc}) are in order.  First, our ``reduced'' theory in
terms of $G(\hat{\bf u})$ can only describe the transitions among the
ferroelectric phases, not the transition to the
high-temperature cubic phase (${\bf u}=0$).  Second, the minimization
over the magnitude of $\bf u$ in Eq.~(\ref{eq:gfunc}) generally
introduces higher orders into the reduced theory.  For example,
truncation of Eq.~(\ref{eq:expan}) at sixth order may still
lead to terms of eighth and higher order in Eqs.~(\ref{eq:gexpan})
and (\ref{eq:fexpan}).  Thus, a statement about what kinds of
critical points can occur simultaneously in an $n^{\rm th}$-order version
of the reduced theory does not necessarily carry over to the
$n^{\rm th}$-order version of the standard Devonshire theory.
Nevertheless, a statement that a certain minimum order of expansion
is needed for the existence of a certain kind of stationary point
in the reduced theory {\it does} carry over to the standard
Devonshire theory.\cite{garcia-private}  For example, we show later
that the reduced theory must be carried to eighth order to allow
for a monoclinic minimum, i.e., an equilibrium $\bf u$ lying in a
(1$\bar1$0) plane.  Then the same statement applies to the standard
Devonshire theory.  For, suppose that $u_0$ is the magnitude of
$\bf u$ at the minimum, and let $G_0(\hat{\bf u})=F(u_0\hat{\bf
u})$; then $G_0$ must have a monoclinic minimum.  But $G_0$ is also
of no higher than eighth order, and so admits only the same type of
minima as does $G$.\cite{garcia-private}

\section{Topological considerations}
\label{sec:topology}

The rules governing the numbers and types of critical or
stationary points of a scalar function defined on a continuous
manifold emerge from a branch of algebraic topology called Morse
theory or, equivalently, the calculus of variations in the
large.\cite{morseetc}  These rules have proved useful in analyzing
lattice vibration spectra\cite{phillips} and the excitation
spectra of crystalline materials generally.  They have not,
however, enjoyed the broad application in condensed matter physics
that might have been expected from these early successes.

The rules take the form of a set of inequalities and one
equality.  The latter is the most powerful, and we confine
ourselves to considering only it explicitly.  $G(\hat{\bf u})$
is an analytic function defined on the surface of a sphere, a
closed two-dimensional manifold of genus zero, for which the
equality becomes
\begin{equation}
N_0-N_1+N_2=2 \;\;,
\label{eq:topolo}
\end{equation}
where $N_0$, $N_1$, and $N_2$ are the number of minima, saddle points,
and maxima, respectively.  This formula assumes only analyticity
in $G(\hat{\bf u})$ and that its Hessian has no vanishing
eigenvalues at the critical points.

Because the function $G(\hat{\bf u})$ has cubic symmetry, any
stationary points that occur must be members of symmetry-related
families of stationary points.  Thus, it is natural to focus
attention on an irreducible wedge corresponding to 1/48 of the
unit sphere, in terms of which Eq.~(\ref{eq:topolo}) can be
rewritten as
\begin{equation}
\sum_j n_j \gamma_j = 2
\label{eq:topol}
\end{equation}
where $j$ runs over the stationary points located in the interior
or on the boundary of the irreducible wedge, $n_j$ is a degeneracy
factor counting the number of images of the stationary point
generated by the cubic symmetry group operations, and $\gamma_j$
is $+1$ for a maximum or minimum and $-1$ for a saddle point.
Symmetry requires only that there be stationary points at T,
O, and R; those at T and R must be maxima or minima while that at
O can be of any type.  Such a symmetry set of stationary points
may or may not be large enough to satisfy the Morse relations
as well.  A set which contains the smallest number
of critical points satisfying both the topological and
symmetry requirements is denoted a minimal set.\cite{phillips}

\begin{figure}
\centerline{\epsfig{file=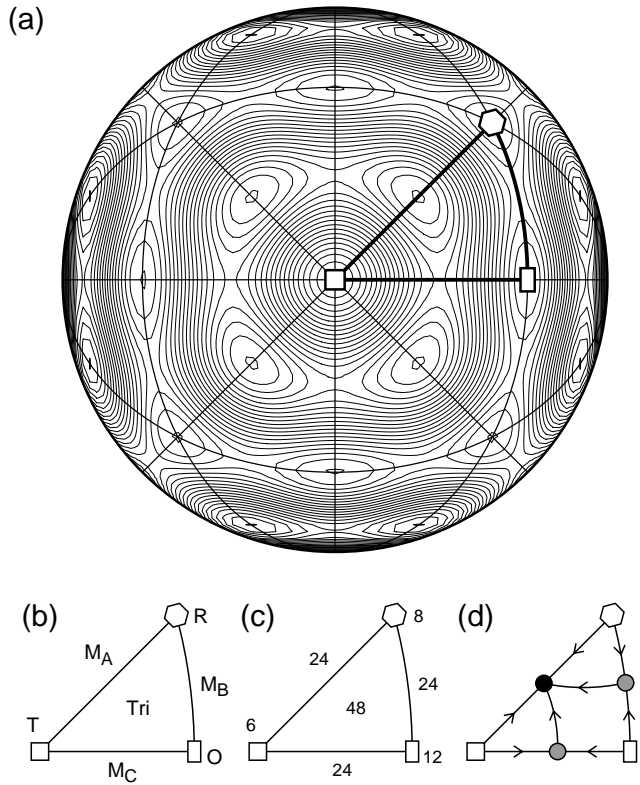,width=3.3in}}
\vspace{0.4cm}
\caption{(a) Contours of constant $G(\hat{\bf u})=-f_8(\hat{\bf u})$
on the unit sphere.
(b) Labels for symmetry points and lines of the irreducible wedge.
(c) Corresponding degeneracy factors $n$ indicating number of
images on the full unit sphere.  (d) Summary representation
of the behavior of $G(\hat{\bf u})$, in which open, shaded, and
filled symbols represent maxima, saddle points, and minima of $G$,
respectively.  Arrows indicate ``downhill'' flow lines.}
\label{fig1}
\end{figure}

This type of analysis is illustrated in Fig.~1.  Figure~1(a) shows
an arbitrarily chosen function $G(\hat{\bf u})$ corresponding to
$c_4=c_6=0$ and $c_8=-1$ in Eq.~(\ref{eq:fexpan}), having maxima
at T, R, and O; saddle points on the lines connecting
T--O and O--R; and a minimum on the segment connecting T--R.
Figure~1(b) specifies the notation that we shall use to identify
the symmetry points and lines.  These are chosen to correspond
to the labels of distorted crystal structures (tetragonal,
rhombohedral, or orthorhombic for the symmetry points; monoclinic
of type `A,' `B,' or `C' for the symmetry lines; and triclinic
for the case of $\hat{\bf u}$ pointing to the interior of the
wedge).  The corresponding degeneracy factors $n$ are given
in Fig.~1(c).  The overall behavior of the function inside the
wedge is summarized in Fig.~1(d), in which the stationary
points are marked by symbols that are open, shaded, or filled
for maxima, saddle points, or minima, respectively.  In this
particular example, it can easily be seen that Eq.~(\ref{eq:topol})
is satisfied (6+8+12$-$24$-$24+24=2).

\begin{figure}
\centerline{\epsfig{file=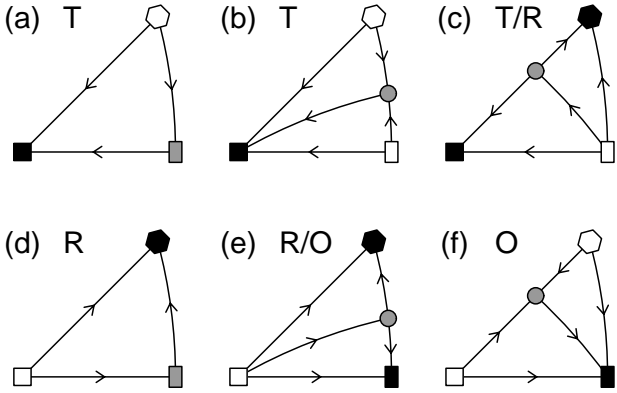,width=3.2in}}
\vspace{0.2cm}
\caption{Sample topologies giving rise to tetragonal (a-c),
rhombohedral (c-e), and orthorhombic (e-f) phases.
Open, shaded, and filled symbols represent maxima, saddle points, and
minima, respectively.}
\label{fig2}
\end{figure}

Using Eq.~(\ref{eq:topol}), one can begin enumerating the
possible topological diagrams for the cubic system.  Figs.~2(a)
and 2(d) show the two possible diagrams in which there are only
three stationary points in the irreducible wedge, the
only possible minimal sets.
In Fig.~2(a) the function has a unique minimum at T, so
the crystal ground state would be tetragonal.  A similar
situation holds for the rhombohedral case of Fig.~2(d).
Figures~2(b-c) and (e-f) show four of the six configurations that
can exist when there are exactly four stationary points in the
wedge.  (Two others, similar to 2(b) and 2(e) but with the the
role reversals T$\leftrightarrow$R and
M$_{\rm B}$$\leftrightarrow$M$_{\rm C}$, are not shown.) In
Fig.~2(c) there are two local minima, so that the ground state
could be of type T or R depending on which minimum is deeper; a
first-order transition between T and R may occur by a crossing
of the minima.  The simplest configuration leading to an
orthorhombic ground state is shown in Fig.~2(f).

In order to arrive at configurations corresponding to monoclinic
or triclinic phases, more than four stationary points are required.
Figs.~3(a-c) illustrate possible configurations for monoclinic
phases of type A, B, and C, in which the order parameter lies
on the T--R, R--O, and O--T lines, respectively.   Figure~3(d)
illustrates a possible triclinic phase.
It is straightforward to check that Eq.~(\ref{eq:topol})
is satisfied for each configuration in Figs.~2 and 3.


Of course, if the expansion of Eq.~(\ref{eq:gexpan}) or (\ref{eq:fexpan})
is truncated at a certain order, the possible types of topological
behavior will be limited by the enforced ``smoothness'' of the
functions allowed at that order.  The purpose of the following
Section is to explore precisely this issue, i.e., to clarify
what types of phases and phase transitions can occur at each
order in the expansion.  With this information in hand, one
can then easily determine what is the minimal model needed to
study a particular physical phenomenon of interest.

\section{Results}
\label{sec:results}

\subsection{Fourth-order theory}
\label{sec:fourth}

If the expansion (\ref{eq:fexpan}) is truncated at fourth
order, the only non-constant cubic invariant is $f_4$ of
Eq.~(\ref{eq:ffuncs}).  The T or R phase is favored as in
Fig. 2(a) or 2(b) if $c_4>0$ or $c_4<0$, respectively.
A transition between T and R phases occurs at $c_4=0$, but this
transition is unphysical because the energy surface is
perfectly flat at the transition.  This degenerate behavior is
an artifact of the truncation to fourth order.

\begin{figure}
\centerline{\epsfig{file=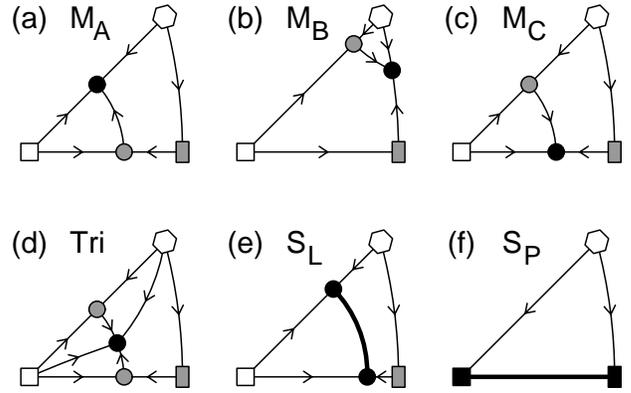,width=3.2in}}
\vspace{0.2cm}
\caption{(a-d) Sample topologies giving rise to monoclinic phases A, B,
and C, and the triclinic phase, respectively.  (e) Degenerate minimum
(heavy solid curve) that can occur when $\beta=3\pi/4$.  (f) Degenerate
minimum (heavy solid line) that occurs for $\alpha=\tan^{-1}(3)$ and
$\beta=\pi$.  Open, shaded, and filled symbols represent maxima, saddle
points, and minima, respectively.}
\label{fig3}
\end{figure}

\subsection{Sixth-order theory}
\label{sec:sixth}

At sixth order in the expansion (\ref{eq:fexpan}), the behavior is
governed by the two coefficients $c_4$ and $c_6$.  Clearly a common
rescaling of the coefficients by a positive scale factor is
irrelevant, so without loss of generality we may set $c_4^2+c_6^2=1$.
It is then convenient to let
\begin{equation}
c_4=\cos(\alpha)\;\;,
\quad 
c_6=\sin(\alpha)\;\;.
\label{eq:alpha}
\end{equation}
By doing numerical calculations and making plots such as that shown
in Fig.~1(a), we find that only three phases exist at this level.
For $-\tan^{-1}(3) < \alpha < \pi/2$, the system is in the T
ground state; for $\pi/2 < \alpha < 5\pi/4$, the system is
in the R ground state; and for $5\pi/4 < \alpha < 2\pi-\tan^{-1}(3)$
the system is in the O ground state.  Fig.~2 illustrates the sequence
of topologies traversed as $\alpha$ is increased.  Starting from
$\alpha=-\tan^{-1}(3)$, one finds T phases corresponding to
Figs.~2(a-c) consecutively, until a first-order transition occurs
to the R state at $\alpha=\pi/2$.  Then one finds R phases
as illustrated in Figs.~2(c-e) until a first-order transition
occurs to the O phase at $\alpha=5\pi/4$.  Finally, O phases
corresponding to Figs.~2(e-f) are found as $\alpha$ is increased
further up to $2\pi-\tan^{-1}(3)$.  The transition from the O to
the T phase is degenerate in the sense that the energy surface
becomes exactly flat along the entire M$_{\rm C}$ symmetry line
at the critical $\alpha$ (this being an artifact of truncation
to low order).

It is important to emphasize that that no monoclinic phase
is possible in the sixth-order model.  In the range
$\pi+\tan^{-1}(3/2) < \alpha < 2\pi-\tan^{-1}(3)$ (i.e.,
$1.3128\pi < \alpha < 1.6024\pi$) in which such a 1D local minimum
appears along the M$_{\rm A}$ symmetry line, it is always unstable
in the second dimension (i.e., it is a saddle point), and the true
minimum is at the O point as in Fig.~2(f).

Further details of the sequence of topologies and the boundaries
between them is given in the Appendix.

\subsection{Eighth-order theory}
\label{sec:eighth}

When the model of Eq.~(\ref{eq:fexpan}) is carried to eighth
order, it no longer becomes profitable to enumerate every possible
topology, as was done above for the sixth-order theory.  Instead,
we choose to focus just on the ``phase diagram'' that is generated
by finding the ground-state symmetry as a function of the parameters
$c_4$, $c_6$, and $c_8$.  A common scaling of the magnitudes of these
coefficients is again unimportant, so we can describe the phase
diagram in terms of two dimensionless parameters that we may take
as
\begin{eqnarray}
c_4&=&\cos(\alpha)\;\;, \nonumber \\
c_6&=&\sin(\alpha)\cos(\beta)\;\;, \nonumber \\
c_8&=&\sin(\alpha)\sin(\beta)\;\;,
\label{eq:ab}
\end{eqnarray}
where $0<\alpha<\pi$ and $0<\beta<2\pi$.  From this point of view,
the ``parameter space'' is just the unit sphere determined by
polar and azimuthal angles $\alpha$ and $\beta$ respectively.

Figure 4 shows the phase diagram that emerges from a careful numerical
study of the minimization of Eq.~(\ref{eq:fexpan}) as a function of
$\alpha$ and $\beta$.  The plot is a mapping
of the unit sphere onto the page.  The points at the ``north''
and ``south'' poles ($\alpha=0$ and $\alpha=\pi$) are the only ones
accessible in the fourth-order theory (Sec.~\ref{sec:fourth});
the dotted vertical lines at $\beta=0$ and $\beta=\pi$ correspond
to the locus of points in parameter space that were explored by
the sixth-order model (Sec.~\ref{sec:sixth}).

\begin{figure}
\centerline{\epsfig{file=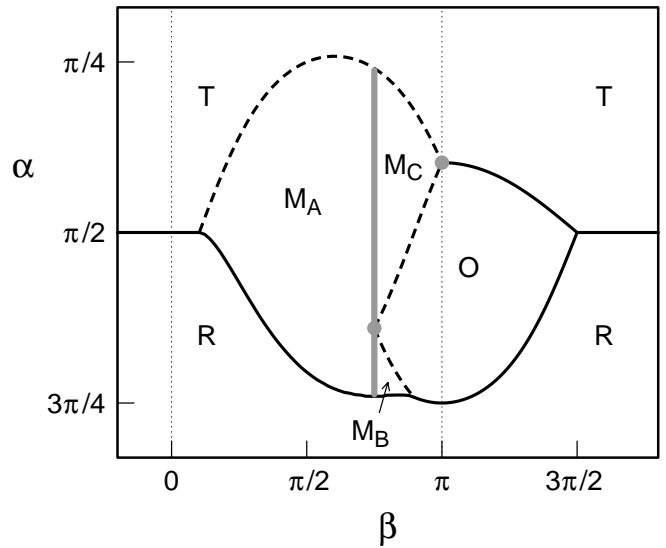,width=3.4in}}
\vspace{0.2cm}
\caption{Phase diagram in the space of parameters $\alpha$ and $\beta$
of the eighth-order theory as defined in Eq.~(\protect\ref{eq:ab}).
Solid and dashed lines are first-order and second-order phase
boundaries, respectively.  The vertical grey line and the grey dots indicate
cases for which degenerate minima occur.  Fine vertical dotted lines
indicate the domain of the sixth-order theory.}
\label{fig4}
\end{figure}

As can be seen from Fig.~4, six of the seven possible phases (i.e.,
possible symmetries of a non-zero order parameter) are accessed by the
eighth-order model.  In addition to the T, R, and O phases that
appeared already at sixth order, all three monoclinic phases
(M$_{\rm A}$, M$_{\rm B}$, and M$_{\rm C}$) are now stable in some region
of the phase diagram.  However, the areas covered by the M$_{\rm C}$
and especially M$_{\rm B}$ regions are relatively small, so these
phases may be harder to find in real systems than the M$_{\rm A}$
phase.

Solid and dashed phase boundaries indicate transitions of first and
second order, respectively, as determined numerically.  According to
Landau theory, there are two necessary conditions for a transition to
be of second order:  (i) the symmetry group of one phase must be a subset
of the symmetry group of the other, ${\cal G}\subset{\cal G}_0$; and
(ii) from the order-parameter displacements that lead from the high-symmetry
to the low-symmetry phase, it should be impossible to construct a 
third-order invariant of ${\cal G}_0$.
\cite{landau}
Transitions of type T--R, T--O, and O--R are necessarily first-order
because of condition (i), and transitions from the R phase to M$_{\rm
A}$ or M$_{\rm B}$ are first-order because of (ii).\cite{explan-third}
As can be seen
from the figure, these are precisely the boundaries that were found to
be of first order.  All others are found to be of second order, with
the exceptions of the M$_{\rm A}$--M$_{\rm B}$ and M$_{\rm A}$--M$_{\rm
C}$ boundaries, which form a vertical line at $\beta=3\pi/4$ indicated
by grey shading in the figure.  Along this line, one finds a degenerate
minimum connecting M$_{\rm A}$ and M$_{\rm C}$ phases as shown in
Fig.~3(e) for $0.259\pi<\alpha<0.640\pi$, and a similar degenerate
minimum connecting M$_{\rm A}$ and M$_{\rm B}$ phases for
$0.640\pi<\alpha<0.740\pi$.  The degenerate behavior can be traced to
the fact that $a_6=0$ along the line $\beta=3\pi/4$.  The triple point
connecting the T, O, and M$_{\rm C}$ phases at $\alpha=\tan^{-1}(1/3)
=0.102\pi$
and $\beta=\pi$ is also a point at which a degenerate minimum occurs,
as shown in Fig.~3(f).  The degenerate behaviors are artifacts of the
eighth-order truncation, as will be explained more fully in
Sec.~\ref{sec:higher}.  Finally, the reader is reminded that
because the theory is based on a single polar orientational order
parameter, Eq.~(\ref{eq:gfunc}), we do not have the ability
to describe transitions to or from the paraelectric C phase, nor
can we describe the more complex AFE or AFD phases.

The variation of some physical variable, such as temperature,
composition, or pressure, will correspond to a variation of the
parameters $\alpha$ and $\beta$ of the model in a way that is
not easy to predict {\it a priori}.  For BaTiO$_3$ and
KNbO$_3$, for which the observed phase transition sequence is
R--O--T--C with increasing temperature, it must be the case
that the system traverses a roughly vertical trajectory on the phase
diagram of Fig.~4, somewhere in the range $\pi<\beta<3\pi/2$.
(Insofar as eighth-order terms are small for these systems,
the trajectory should be near $\beta=\pi$.) 
The R--O--T sequence can be visualized as traversing Figs.~2(d-f)
and 2(a), in that order.

On the other hand, systems such as PZT\break (PbZr$_{1-x}$Ti$_x$O$_3$)
that exhibit a morphotropic phase boundary (MPB), i.e., an R--T
transition as a function of composition $x$, 
evidently cross the first-order R--T phase boundary in the
vicinity of $\beta=0$ with increasing $x$.  If this trajectory
passes to the right of the triple point connecting R, T, and
M$_{\rm A}$ phases at $\alpha=\pi/2$, $\beta=\tan^{-1}(1/3)=0.102\pi$ in
Fig.~4, then the phase transition sequence becomes R--M$_{\rm A}$--T,
as recently observed experimentally.\cite{noh-apl,noh-prb,noh-prb2}
The narrowness of the range of M$_{\rm A}$ phase, only a few
percent in $x$, suggests that the trajectory passes rather close to the
triple point.  In fact, there are strong experimental indications
of the possible existence of a triple point in the $x$--$T$ phase
diagram of PZT near $x\simeq0.47$ and $T\simeq100^\circ$C.
\cite{noh-apl,noh-prb,noh-prb2}  Thus, it may be that the behavior near the
triple point can be explored experimentally in the PZT system.

Using the topological analysis introduced earlier, we can now
clarify the nature of the T--M$_{\rm A}$--R transition sequence
near the triple point.  Referring to Fig.~5, we imagine traversing
a downward trajectory of increasing $\alpha$ at fixed $\beta$,
slightly to the right of the triple point.  Starting deep in the
T phase, Fig.~2(a), a saddle point detaches from the O point and
traverses the M$_{\rm B}$ line toward the R point, Fig.~2(b).
After it passes through the R point and emerges on the ``other
side'' (on the M$_{\rm A}$ line), we find ourselves in the situation
of Fig.~5(a), the R point having been converted to a local minimum.
Up to this point, the global minimum remains at T.  Next, the T
point converts from a local minimum to a local maximum, with the
simultaneous emission of a saddle point along M$_{\rm C}$ and a
local minimum along M$_{\rm A}$, as shown in Fig.~5(b).  This event
corresponds to the second-order T--M$_{\rm A}$ transition.  The
first-order M$_{\rm A}$--R transition then occurs by the crossing
of the energies of the local minima of Fig.~5(b).  Once in the R
phase, the local minimum and saddle point on the M$_{\rm A}$ line
annihilate one another to give Fig.~5(c).  Finally, deep in the R
phase, the M$_{\rm C}$ saddle point eventually arrives at O, giving
rise to the situation of Fig.~2(d).

\begin{figure}
\centerline{\epsfig{file=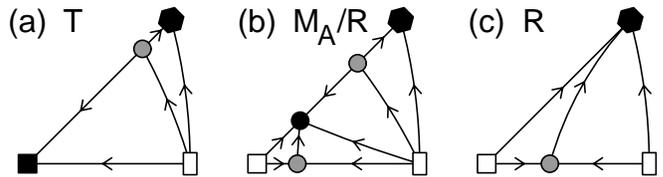,width=3.4in}}
\vspace{0.2cm}
\caption{Topologies encountered in the T--M$_{\rm A}$--R transition
sequence near the triple point.  (a) Topology of T phase near
the second-order T--M$_{\rm A}$ transition.  (b) Topology of M$_{\rm A}$
phase and, as well, of R phase near the first-order M$_{\rm A}$--R transition.
(c) Topology occurring deeper in the R phase.}
\label{fig5}
\end{figure}

It should thus be emphasized that the present theory makes a definite
prediction about the nature of the transitions that occur in
the T--M$_{\rm A}$--R transition sequence.  We can predict that,
if it were possible to scan with decreasing $x$ at a temperature
below that of the triple point, one would first find a continuous
rotation of the polarization from [001] into the (1$\bar1$0) plane
starting at a critical $x_{c2}$, and then a discontinuous jump to
the [111] direction when there is a crossing of the free energies
of the M$_{\rm A}$ and R phases at $x_{c1}$.  Unfortunately, the
fact that the T--M$_{\rm A}$ and especially the R--M$_{\rm A}$
boundaries lie almost vertically in the experimental $x$--$T$ plane
may make it difficult to test this prediction, since $x$ can only
be varied by preparation of multiple samples. Nevertheless, this
scenario seems to be supported by the numerical simulations of
Ref.~\onlinecite{bellaiche}.

It is important to note that a triclinic phase does not occur
anywhere in the phase diagram of the eighth-order model.
However, we do note the possibility of observing new monoclinic
phases of types M$_{\rm B}$ and M$_{\rm C}$ in a region near
$\beta\simeq0.8\pi$.  In fact, the sixth-order model (vertical
dotted line at $\beta=\pi$) comes very close to yielding
a monoclinic M$_{\rm C}$ phase near the triple point (grey dot
at $\alpha$=$\tan^{-1}(3)$, $\beta$=$\pi$) where the M$_{\rm C}$,
T, and O phases are in equilibrium.  If a system such as
BaTiO$_3$ or KNbO$_3$ could somehow be perturbed so that the
variation with temperature would carry the system on a
trajectory passing to the left of this triple point in
Fig.~4, then a novel R--O--M$_{\rm A}$--T--C (or even
R--M$_{\rm B}$--O--M$_{\rm A}$--T--C) transition
sequence might be observed.  However, to our knowledge, no
M$_{\rm B}$ or M$_{\rm C}$ phase has ever been observed in
a cubic perovskite system.

\subsection{Higher-order expansions}
\label{sec:higher}

We have seen that the eighth-order expansion still does not allow
for the appearance of a triclinic equilibrium phase for any
parameter values.  It is natural, then, to ask at what higher order
in the expansion a triclinic phase can first occur.  The answer
is that the expansion must be carried to {\it twelfth order}
before a triclinic phase can appear.

For, suppose that a triclinic phase is the ground state.  Then the
minimum of $G$ occurs at a point $\hat{\bf u}_0$ located in the
interior of the irreducible wedge, as illustrated in Fig.~3(d).
Letting
\begin{eqnarray}
\widetilde{G}(\hat{\bf u}) &=& G(\hat{\bf u}) 
                          -G(\hat{\bf u}_0) \;\;, \nonumber\\
\tilde{g}_4(\hat{\bf u}) &=& g_4(\hat{\bf u})
                            -g_4(\hat{\bf u}_0) \;\;, \nonumber\\
\tilde{g}_6(\hat{\bf u}) &=& g_6(\hat{\bf u})
                            -g_6(\hat{\bf u}_0) \;\;,
\label{eq:gtilde}
\end{eqnarray}
the expansion Eq.~(\ref{eq:gexpan}) can be rewritten
\begin{equation}
\widetilde{G}(\hat{\bf u}) =
  \tilde{a}_8 \tilde{g}_4^2
+ \tilde{a}_{10} \tilde{g}_4 \tilde{g}_6
+ \tilde{a}_{12} \tilde{g}_4^3
+ \tilde{a}_{12}' \tilde{g}_6^2
+ ... \;\;,
\label{eq:expant}
\end{equation}
where the $\tilde{a}_n$ are trivial linear combinations of the $a_n$.
To drop the $\tilde{a}_4$ and $\tilde{a}_6$ terms, we have used the
fact that $G(\hat{\bf u})$ must be stationary at $\hat{\bf u}_0$.  (For
this we also need that the gradients of $g_4$ and $g_6$ never vanish or
become parallel at an interior point of the irreducible wedge; this
is straightforward to confirm.)

It is now evident that if expansion (\ref{eq:expant}) is truncated at
eighth order, then point $\hat{\bf u}_0$ is not an isolated minimum.
Instead, it belongs to a degenerate locus of minima corresponding to
$\tilde{g}_4=0$, i.e., to a contour of the function $g_4(\hat{\bf u})$.
This is the situation illustrated in Fig.~3(e).
It occurs when $a_6=0$, i.e., when $c_6+c_8=0$, corresponding
to the grey vertical line at $\beta=3\pi/4$ in Fig.~4.

If the expansion is carried to tenth order, then it is clear from
Eq.~(\ref{eq:expant}) that $\widetilde{G}$ still
vanishes on this same contour.  The two-dimensional Hessian matrix
$H_{\mu\nu}=d^2 G/ d\hat{u}_\mu d\hat{u}_\nu$ then takes the form
\begin{equation}
H = \pmatrix{ 0 & d \cr d & e \cr}  \;\;,
\label{eq:hessian}
\end{equation}
where indices $\mu$=1 and 2 correspond to the directions parallel and
perpendicular to the $g_4$ contour, respectively. $d\ne0$ as long as 
$\tilde{a}_{10}\ne0$, in which case $\det H<0$.
Thus, at tenth order, the stationary point
$\hat{\bf u}_0$ cannot be a minimum; instead, it is generally an
isolated saddle point.

Finally, it is clear that the point $\hat{\bf u}_0$ can be a local
minimum if the expansion is carried to twelfth order.  For suppose
$\tilde{a}_{10}$=$\tilde{a}_{12}$=0, and $\tilde{a}_8$ and
$\tilde{a}_{12}'$ are positive.  Then $\widetilde{G}=
\tilde{a}_8 \tilde{g}_4^2+\tilde{a}_{12}' \tilde{g}_6^2$ is obviously
positive definite, and $\hat{\bf u}_0$ is a true isolated (global) minimum.

Concluding this section, we find that it is necessary to go to
surprisingly high order in the free-energy expansion in order to
stabilize a ferroelectric state in which there are no symmetry
constraints on the order parameter $\bf P$.  Specifically,
we find that cubic invariants of {\it twelfth} or higher order have
to be included to to stabilize such a triclinic phase.  We conclude
that the discovery (or synthesis) of a material having such behavior
may be challenging, but is by no means impossible.

\section{A microscopic model}
\label{sec:model}

When structural transitions have some order-disorder character,
a model free energy (e.g., (\ref{eq:expan}) or (\ref{eq:fexpan}))
expressed as a function of a macroscopic order parameter (e.g.,
$\bf u$ or $\hat{\bf u}$) provides little insight into the
local structural fluctuations that underlie the transitions.
In such a case, a more appropriate microscopic picture of the
high-symmetry phase may be one in which local regions have undergone
a symmetry-lowering structural distortion, but in such a way that
long-range order has not set in.  For example, the description
of the phase transition sequence of BaTiO$_3$ and KNbO$_3$ in
terms of the well-known ``eight-site model''\cite{eight-site}
assumes the presence of random local rhombohedral displacements
in the orthorhombic, tetragonal, and cubic phases.  In the
present case, Noheda {\it et al.} have concluded from their own
structural analysis of PZT\cite{noh-apl,noh-prb,noh-prb2} and that of
Corker {\it et al.}\ of rhombohedral PZT,\cite{corker} that there
may be random local {\it monoclinic} displacements which order
variously to yield the tetrahedral, rhombohedral, or monoclinic
phases near the morphotropic phase boundary.   One could then
describe the system in terms of fluctuations between minima of a
``24-site model.''

To make these ideas more precise, suppose that the local
displacements of Noheda {\it et al.} arise from an optical branch
of the phonon spectrum, and let ${\bf u}_l$ be the vector
``local mode amplitude'' for that branch within the $l$th unit
cell.\cite{zhong}  Take as a model of the free energy
\begin{equation}
F[{\bf u}] = \sum_l f({\bf u}_l)\,+\,{1 \over 2}
{\sum_{lm}}' f_{lm}({\bf u}_l,{\bf u}_m) \;\;.
\label{eq:fmodel}
\end{equation}
The in-cell energy $f$ is presumed to be strongly nonlinear;
the inter-cell coupling $f_{lm}$ may either be bilinear in
${\bf u}_l$ and ${\bf u}_m$, or of higher order.  $f$ is also
presumed to dominate the $f_{lm}$ so that, to a good
approximation, its global minima establish the possible magnitude
and orientations ${\bf u}_l^\alpha$ of ${\bf u}_l$.
$f({\bf u}_l)$ can be subjected to precisely the same methods of
analysis as applied to $F({\bf u})$ in
Secs.~\ref{sec:form}-\ref{sec:results} above, with parallel
results.  The ${\bf u}_l^\alpha$ so obtained can then be
substituted into $F$ in Eq.~(\ref{eq:fmodel}), yielding.
\begin{equation}
F[\alpha] = F_0 \,+\,{1 \over 2}
{\sum_{lm}}' f_{lm}({\bf u}_l^\alpha,{\bf u}_m^\beta) \;\;.
\label{eq:potts}
\end{equation}
This can be regarded as a 24-site version of the $q$-state Potts model,

A statistical analysis can be carried out for various forms of
$f_{lm}$ (e.g., bilinear) to capture the cubic (completely
disordered), tetragonal (partially disordered), rhombohedral
(partially disordered), and monoclinic (fully ordered) phases of
the case where ${\bf u}_l^\alpha$ takes on all of the 24
symmetrically equivalent monoclinic displacements.  This type of
analysis has already been carried out for orientational
order-disorder transitions and plastic crystals.\cite{pick}

\section{Summary and conclusions}
\label{sec:sum}

The original Devonshire theory\cite{dev} gave a natural explanation
for the appearance of tetragonal, orthorhombic, and rhombohedral
phases in materials such as BaTiO$_3$ based on a sixth-order
free-energy expansion.  Here, we have confirmed that these
ferroelectric phases, in which the order parameter $\bf P$ is
confined to a symmetry axis, are the only ones permitted by the
sixth-order version of the theory.  Moreover, we have clarified
the nature of the phases that may be expected to appear at higher
orders in the expansion.  In particular, we have shown that the
extension of the theory to eighth order allows one to describe, in
addition, three kinds of monoclinic phase in which $\bf P$ is confined
only to a symmetry plane.  To obtain a triclinic phase
in which $\bf P$ is unconstrained by symmetry, we have shown that
a twelfth-order version of the theory is needed.
A topological analysis of the critical points of the energy surface
has been used to facilitate the discussion of the relevant phases and
phase transitions.

The present theory may provide some added insight into the phase
behavior of conventional ferroelectrics such as BaTiO$_3$, but
the principal new results concern cases in which the
eighth-order terms are important.  In particular, the theory
provides a natural explanation for the monoclinic M$_{\rm A}$
phase recently observed experimentally in PZT.\cite{noh-apl,noh-prb,noh-prb2}
It also predicts that if a triple point of equilibrium
between T, R, and M$_{\rm A}$ phases occurs, then it will be one
at which first-order R--T and R--M$_{\rm A}$ boundaries meet a
second-order T--M$_{\rm A}$ boundary.  That is, $\bf P$ will
rotate into the (1$\bar1$0) mirror plane continuously from the
T side but discontinuously from the R side.

Noheda {\it et al.}\cite{noh-prb2} have shown such a triple point
in their Fig.~6, in agreement with our analysis.  However, they
also show a region of coexistence of the T and M$_{\rm A}$ phases
in the same phase diagram.  In a homogeneous sample, such a
coexistence region can be due to hysteresis arising from
nucleation barriers to a first-order phase transition.  The
eighth-order theory predicts the T--M$_{\rm A}$ phase boundary to
be of second order, in which case there can be no hysteresis or
nucleation barriers.  In principle, higher-order contributions
to the free energy could be large enough to change the order of
the T--M$_{\rm A}$ transition.  However, a more likely
explanation is that the samples studied by Noheda {\it et al.}
may be inhomogeneous.  We note that these authors did not report
hysteresis but did report a two-phase coexistence region near
the R--T boundary between 500 and 575\,K in Fig.~5 of
Ref.~\onlinecite{noh-prb2}  Attributing this to inhomogeneity, we
estimate that the concentration variation may be of order 1\% from
the slope of the R--T/M--T phase boundary in Fig.~6 of
Ref.~\onlinecite{noh-prb2}.  Composition inhomogeneity of
that magnitude would be sufficient to account for the T--M$_{\rm
A}$phase coexistence shown in that figure.

Finally, the work may provide some guidance in the search for
even more novel M$_{\rm C}$, M$_{\rm B}$, and triclinic
ferroelectric phases.  It also may be of utility in other kinds
of cubic systems with other kinds of vector order parameters,
e.g., ferromagnetic systems.

\acknowledgments

D.V.~acknowledges support of ONR Grant N00014-97-1-0048.
We wish to thank L.~Bellaiche, A.~Garc\'{\i}a, A.~Khachaturyan, and
B.~Noheda for useful comments and discussions.

\appendix
\section{Details of sixth-order theory}
\label{sec:app}

The purpose of this Appendix is to give further details about
the sequence of transitions that occurs in the sixth-order
model of Sec.~\ref{sec:sixth}.  Recall that the behavior in
this model is governed by a single dimensionless parameter
$\alpha$ defined via Eq.~(\ref{eq:alpha}).

There are ten critical values of $\alpha$ that we can define
as
\begin{eqnarray}
\alpha_1   &=&-0.3976\pi=-\tan^{-1}(3)    \nonumber\\
\alpha_2   &=& 0.2256\pi= \tan^{-1}(6/7)  \nonumber\\
\alpha_3   &=& 0.25  \pi                  \nonumber\\
\alpha_4   &=& 0.3128\pi= \tan^{-1}(3/2)  \nonumber\\
\alpha_5   &=& 0.5   \pi                  \nonumber\\
\alpha_6   &=& 0.6024\pi=\pi-\tan^{-1}(3)    \nonumber\\
\alpha_7   &=& 1.2256\pi=\pi+\tan^{-1}(6/7)  \nonumber\\
\alpha_8   &=& 1.25  \pi                  \nonumber\\
\alpha_9   &=& 1.3128\pi=\pi+\tan^{-1}(3/2)  \nonumber\\
\alpha_{10}&=& 1.5   \pi
\label{list}
\end{eqnarray}
The sequence of phases can be followed on Fig.~4 by tracing
the vertical dotted lines, first from top to bottom at
$\beta=0$ for $0<\alpha<\pi$, and then from bottom to top
at $\beta=\pi$ for $\pi<\alpha<2\pi$.

In the T phase, the system exhibits the topology of
Fig.~2(a) for $\alpha_1<\alpha<\alpha_2$;
Fig.~2(b) for $\alpha_2<\alpha<\alpha_4$;
and Fig.~2(c) for $\alpha_4<\alpha<\alpha_5$.
At $\alpha_2$, point O converts to a local maximum and simultaneously
a saddle point appears along the M$_{\rm B}$ symmetry line.
There is an irrelevant crossing of the R and O maxima at $\alpha_3$.
At $\alpha_4$, R becomes a local minimum, and the saddle point
switches from the M$_{\rm B}$ to the M$_{\rm A}$ symmetry line.
The transition from T to R is first order at $\alpha_5$.

The system falls into an R ground state corresponding to
Fig.~2(c) for $\alpha_5<\alpha<\alpha_6$;
Fig.~2(d) for $\alpha_6<\alpha<\alpha_7$;
and Fig.~2(e) for $\alpha_7<\alpha<\alpha_8$.
The M$_{\rm A}$ saddle point disappears and T is converted
to a maximum at $\alpha_6$, and a new M$_{\rm B}$ saddle point
emerges with the conversion of O to a local minimum
at $\alpha_7$.  The transition from R to O is first order at
$\alpha_8$.

Finally, an O phases occurs as illustrated in Fig.~2(e) for
$\alpha_8<\alpha<\alpha_9$, and as in Fig.~2(f) for
$\alpha_9<\alpha<\alpha_1+2\pi$.  The saddle point at M$_{\rm B}$
vanishes and R is converted to a local maximum at $\alpha_9$.
There is an irrelevant crossing of the T and R maxima at
$\alpha_{10}$.  The transition from O to T at $1.6024\pi$ is
singular, in that the energy surface becomes exactly flat along
the entire M$_{\rm C}$ symmetry line.



\begin{references}
\bibitem{explan-third}
Representing $\hat{\bf u}$ by polar coordinates $(\theta,\phi)$
measured with respect to the (111) axis, a function such as
$\sin^3(\theta)\cos(3\phi)$ is third order in displacements
from $\theta=0$ and is invariant with respect to the rhombohedral
symmetry group.
\end{references}
\end{document}